\documentclass[global,twocolumn,final]{svjour}	

\usepackage{graphics}
\usepackage{cite}
\usepackage{verbatim}
\usepackage{amsmath}
\usepackage{amssymb}
\usepackage{amsfonts}

\journalname{}

\begin{document}

\title{Microwave control electrodes for scalable, parallel, single-qubit operations in a surface-electrode ion trap}

\author{D. P. L. Aude Craik \and N. M. Linke  \and T. P. Harty  \and C. J. Ballance  \and D. M. Lucas  \and A. M. Steane  \and D. T. C. Allcock\thanks{Present address: National Institute of Standards and Technology, 325 Broadway, Boulder, Colorado 80305, USA} }

\institute{Department of Physics, University of Oxford, Clarendon Laboratory, Parks Road, Oxford, OX1 3PU, UK \email{d.pradolopesaude@physics.ox.ac.uk}}

\date{Received: date / Revised version: date} 

\maketitle

\begin{abstract}
We propose a surface ion trap design incorporating microwave control electrodes for near-field single-qubit control. The electrodes are arranged so as to provide arbitrary frequency, amplitude and polarization control of the microwave field in one trap zone, while a similar set of electrodes is used to null the residual microwave field in a neighbouring zone. The geometry is chosen to reduce the residual field to the 0.5\% level without nulling fields; with nulling, the crosstalk may be kept close to the 0.01\% level for realistic microwave amplitude and phase drift. Using standard photolithography and electroplating techniques, we have fabricated a proof-of-principle electrode array with two trapping zones. We discuss requirements for the microwave drive system and prospects for scalability to a large two-dimensional trap array.
\end{abstract}

\section{Introduction}

Trapped-ion hyperfine ground level atomic ``clock" qubits are promising candidates for building a quantum computer that operates below the fault-tolerant threshold.  The advantageous features of these qubits are practically infinite $T_{1}$ lifetimes, qubit transitions with frequencies in the convenient few-GHz regime and lack of first-order Zeeman shifts (magnetic field fluctuations are the main cause of dephasing in trapped-ion qubits).

Long coherence times and high-fidelity single-qubit state preparation, gates and readout have recently been simultaneously demonstrated at a level sufficient to support fault-tolerant quantum computing in a $^{43}$Ca$^+$ qubit \cite{Harty13}.  Comparable single-qubit gate fidelities and coherence times have also been demonstrated in a $^9$Be$^+$ qubit \cite{Langer2005, Bro11}. Two-qubit gates have been implemented using microwaves in $^{25}$Mg$^+$ qubits \cite{Osp11}. These results were all achieved using incoherent optical processes for the preparation and readout steps and direct microwave manipulation of the qubit for coherent operations. The microwave-driven operations were carried out in the near-field regime with the ions tens of microns above microwave conductors.

The near-field regime offers several important advantages over the free-space microwave and laser-based control techniques that are more commonly used. Firstly, compared to the free-space regime, the near-field regime gives stronger coupling. This allows for fast  single-qubit gates (sub-$\mu$s) at reduced powers and avoids the non-linearities and transient thermal effects associated with power amplifiers.  Secondly, the coupling strength and polarization at the ion are very stable as the microwave conductors are registered to the trap structure. Thirdly, the combination of several conductors in the near field allows for relatively simple polarization control of the microwave field \cite{Sha13}.  This enables efficient coupling to the qubit transition and suppression of unwanted light-shifts and off-resonant excitation due to other transitions within the hyperfine manifold; this in turn permits fast gates with Rabi frequencies that are not small compared to the hyperfine splitting.  Finally, in the near field regime, the microwave field has a strong spatial dependence which enables the addressing techniques discussed in this paper and facilitates the state-dependent motional transitions required in multi-qubit gates \cite{Osp08}.

There are two types of laser-driven gate in common use: two-photon Raman transitions on ground-level qubits and quadrupole transitions on optical qubits.  Raman transitions suffer from  photon scattering errors \cite{Ozeri07}.  Optical qubits suffer from the finite $T_{1}$ lifetime of the upper qubit state and from the fact that it is difficult to stabilize a laser's frequency and phase to the same absolute stability achievable with a microwave source.  Both suffer from laser beam pointing noise.

The current-carrying microwave electrodes required to drive gates are straightforward to integrate into present microfabricated trap designs as these already feature the large number of electrodes ($\sim 100$) necessary for generating the trapping fields.  This is a significant simplification compared to the prospect of integrating hundreds of laser beams into a large-scale trapped-ion processor, for example via integrated optics \cite{Str11}.

Laser access will still be required but is limited to cooling, preparation and readout steps.  These processes  require very low power compared to coherent optical operations (typically ${\rm\mu }$Ws rather than mWs) and are more robust to amplitude, phase, frequency and polarization noise. These parameters are difficult to control in integrated optics \cite{Brady10}, and limiting the use of lasers to robust processes may be critical to making optical integration feasible.

There are two major technical hurdles associated with near-field microwave techniques. The first is the fidelity of the two-qubit gates.  This was limited to an error per gate of 0.24 in the single demonstration experiment performed so far\cite{Osp11}, though, with technical improvements, it appears this could be reduced significantly \cite{War13}.  This paper will address the second limitation: crosstalk. Crosstalk occurs because a microwave current applied to an electrode produces a finite microwave field across the entire processor, leading to  unwanted operations on qubits other than the target qubit.  This is in contrast to a focused laser beam, where the intensity falls away to a negligibly small value only a few beam waists away from the target ion.

It has recently been experimentally demonstrated that, despite this crosstalk, single-ion addressing can be performed with high fidelity (Rabi frequency ratio below 2\% for a pair of ions in a two ion string)\cite{War12}. However, this implementation is limited in that it uses large transverse microwave gradients that are only suitable for a linear array of ions, and requires very precise and stable positioning of the ions.  The radial ion displacements used as the addressing mechanism also introduce micromotion, which can be undesirable.  Finally, although differences in microwave field amplitude can be produced at each ion, differences in phase cannot and polarization control is limited.  In this paper, we propose a scheme to produce arbitrary microwave amplitude, phase and polarization at each location on a two-dimensional array of ions using smaller gradients and without introducing micromotion.

Single-ion addressing and a two-qubit gate using microwaves have also been achieved using a static magnetic field gradient to create a differing Zeeman shift at different ions \cite{Kun13, Joh09, Wang09}. Compared with near-field techniques, this method uses a simpler trap design, but the processing bandwidth (i.e. the number of single qubit operations per unit time) is limited by the size of the Zeeman splitting that can be generated, the method requires field-dependent states (which have short coherence times) and, since each qubit has a different frequency, tracking of all the qubit phases during a computation may be challenging.

\section{Principle of operation}

We propose a large ion trap array with multiple trapping zones (as demonstrated in \cite{Ami10}, for example).  A static magnetic field $\bf B_0$ defines a quantization axis.  Each trapping zone stores a single ion in close proximity to a set of microwave control electrodes.  These electrodes guide currents which generate a microwave near-field at the ion.  Each electrode is driven at the qubit frequency, with independently adjustable phase and amplitude.  If at least three control electrodes are provided for each ion, there are enough degrees of freedom to produce a microwave field of any (including zero) amplitude, phase and polarization at each ion.  In what follows, we will consider a trap with four electrodes per ion, as this provides for a symmetric trap design.

The control electrodes will also create finite fields (crosstalk) at the other trapping locations, both directly and via induced currents in other electrodes.  The fields generated at each ion can be described by the equation

\begin{equation}
\label{fieldeqn}
\left(\begin{array}{c}
	\mathbf{B}_{1,x}\\
	\mathbf{B}_{1,y}\\
	\mathbf{B}_{1,z}\\
	\vdots\\
	\mathbf{B}_{N,x}\\
	\mathbf{B}_{N,y}\\
	\mathbf{B}_{N,z}
	\end{array}\right)
	=
\mathbf{M}
\times
\left(\begin{array}{c}
	\mathbf{I}_{1,a}\\
	\mathbf{I}_{1,b}\\
	\mathbf{I}_{1,c}\\
	\mathbf{I}_{1,d}\\
	\vdots\\
	\mathbf{I}_{N,a}\\
	\mathbf{I}_{N,b}\\
	\mathbf{I}_{N,c}\\
	\mathbf{I}_{N,d}
\end{array}\right)
\end{equation}
where $\mathbf{B}_{n,x}=B_{n,x}e^{i\phi_{n,x}}$ is the $x$-component of the microwave B-field at ion $n = 1 \ldots N$.  We have labelled the 4 electrodes around each ion $a \ldots d$ and $\mathbf{I}_{n,a}=I_{n,a}e^{i\phi_{n,a}}$ is the current applied to electrode $a$ around ion $n$.  $\mathbf{M}$ is a $3N\times4N$ matrix that describes the couplings and crosstalk between each control electrode and each ion.

In practice, $\mathbf{M}$ must be determined experimentally.  The amplitude of each matrix element can be found by applying a microwave signal to one control electrode at a time and measuring Rabi frequencies (for the spatial component of $\mathbf{B}$ which drives the qubit transition) or light shifts (for the components of $\mathbf{B}$ which drive other off-resonant transitions) at each ion.  The relative phases of the elements can be determined by running the control electrodes pairwise and observing the change in Rabi frequency or light shift as a function of relative phase.

Once $\mathbf{M}$ is known, a set of control currents can be calculated that will produce the required $\mathbf{B}$-field to perform any single-qubit operation at any ion.  Since the effects of crosstalk are included in this calculation, it in principle introduces no error to the precision with which $\mathbf{M}$ can be measured and $\mathbf{B}$ can be set, provided the system is linear.  This is similar to the method used to generate electrical potentials of the desired shape in ion traps by using a superposition of many electrode contributions \cite{Blakestad10, Allcock10}. Fig.~\ref{nullschematic} shows schematically how, for a two-zone trap, an operation on ion 1 can be performed whilst performing the identity on ion 2. 

\section{Experimental Design}

We propose a proof-of-principle experiment using two of the microwave addressing zones described above.

\subsection{Qubit}

We use ground-level hyperfine states in $^{43}$Ca$^+$ ions as our qubit.  At low static fields of $B_0 \lesssim $10\,G, the 3.2\,GHz $S^{4, 0}_{1/2}\leftrightarrow S^{3, 0}_{1/2}$ transition (where the superscript indicates the angular momentum quantum numbers $F$, $M_F$) has only a small first-order magnetic field dependency and has been shown to be a robust qubit \cite{Lucas2007}.  We can also use intermediate-field clock qubits (for example $S^{4, 0}_{1/2}\leftrightarrow S^{3, +1}_{1/2}$ at $B_0$=146\,G or $S^{4, +1}_{1/2}\leftrightarrow S^{3, +1}_{1/2}$ at $B_0$=288\,G), which have zero first-order magnetic field dependency. Both the low field and 146\,G qubits have already been used for experiments involving near-field microwave manipulation \cite{Allcock13, Harty13}.

\subsection{Trap design}

The trap was designed and simulated using Ansoft HFSS, a finite-element electromagnetic simulation software package.  HFSS allows us to calculate the current distributions in the electrodes (see fig.~\ref{hfss}) and the microwave near-fields generated above them.

\begin{figure}
\begin{center}
\resizebox{0.48\textwidth}{!}{\includegraphics{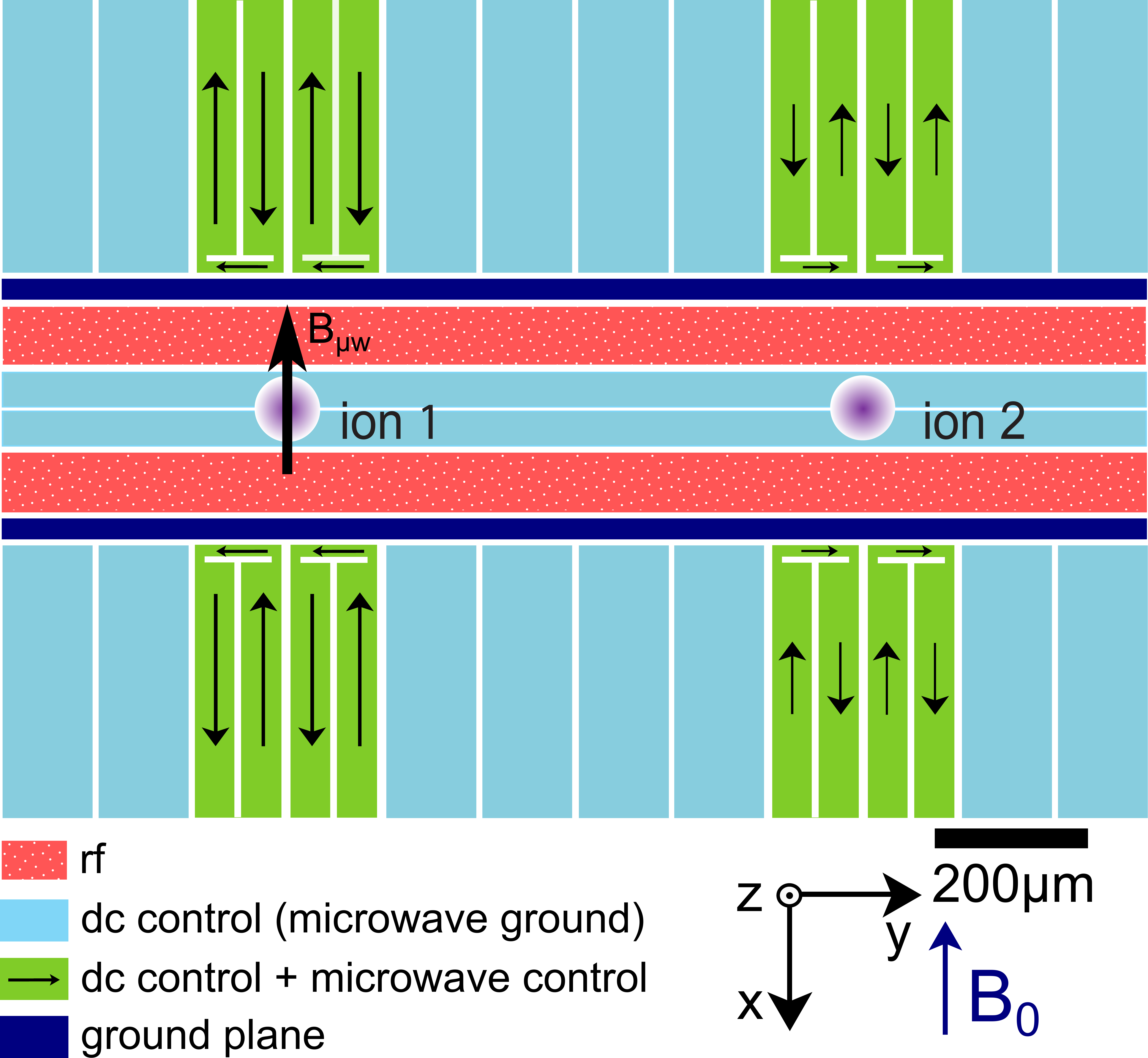}}
\end{center}
\caption{Schematic of our proposed proof-of-principle trap.  In the example shown, we can address ion 1 without crosstalk at ion 2 because the crosstalk is nulled by the electrodes at ion 2.}
\label{nullschematic} 
\end{figure}

\subsubsection{Trapping zone}

The trap is a linear surface-electrode trap \cite{Chi05} with all electrodes symmetric about the trap's centre (see fig. \ref{nullschematic}).  The split axial DC electrode allows for orientation of the radial modes as required for optimal Doppler cooling \cite{Allcock10}.  The RF electrodes are 101.5\,$\mu$m wide and the axial DC electrodes are 60\,$\mu$m wide: their dimensions were chosen using the theory developed in \cite{Hou08, Wesenberg08}. The DC/microwave control electrodes are 150\,$\mu$m wide and the ground plane between the RF and DC/micro\- wave control electrodes is 35\,$\mu$m wide.  All gaps are 10\,$\mu$m wide, except for the central gap, which is 5\,$\mu$m wide.  The ion to surface distance is calculated to be 110\,$\mu$m.  The RF electrodes were made as narrow as possible without significantly compromising trapping performance in order to to minimize the distance between the microwave control electrodes and the ions.

\subsubsection{Microwave control electrodes}

The microwave control electrodes are created by adding a T-shaped slot to a DC electrode so that it forms a current loop capable of generating an oscillating magnetic field at the ion when microwave currents are applied.  This current is generated by connecting one end of the loop to a microwave source and shorting the other end to ground via a capacitor (fig. ~\ref{package}b).  Since the electrode length is short ($\sim$4.5\,mm) compared to $\lambda/4$ (15.1\,mm for 3.2\,GHz using a fused silica substrate with an effective dielectric constant of $\epsilon_{r}=2.41$), the whole loop is close to the current anti-node of the generated standing wave.   The electrodes must still fulfil the requirements of a standard DC control electrode, i.e. they must act as an RF ground and their DC voltage must be controllable.  The former is achieved via the grounding capacitor and the latter by using a bias-T to add a DC voltage to the microwave input.

\subsubsection{Chip layout}

\begin{figure}
\begin{center}
\resizebox{0.48\textwidth}{!}{\includegraphics{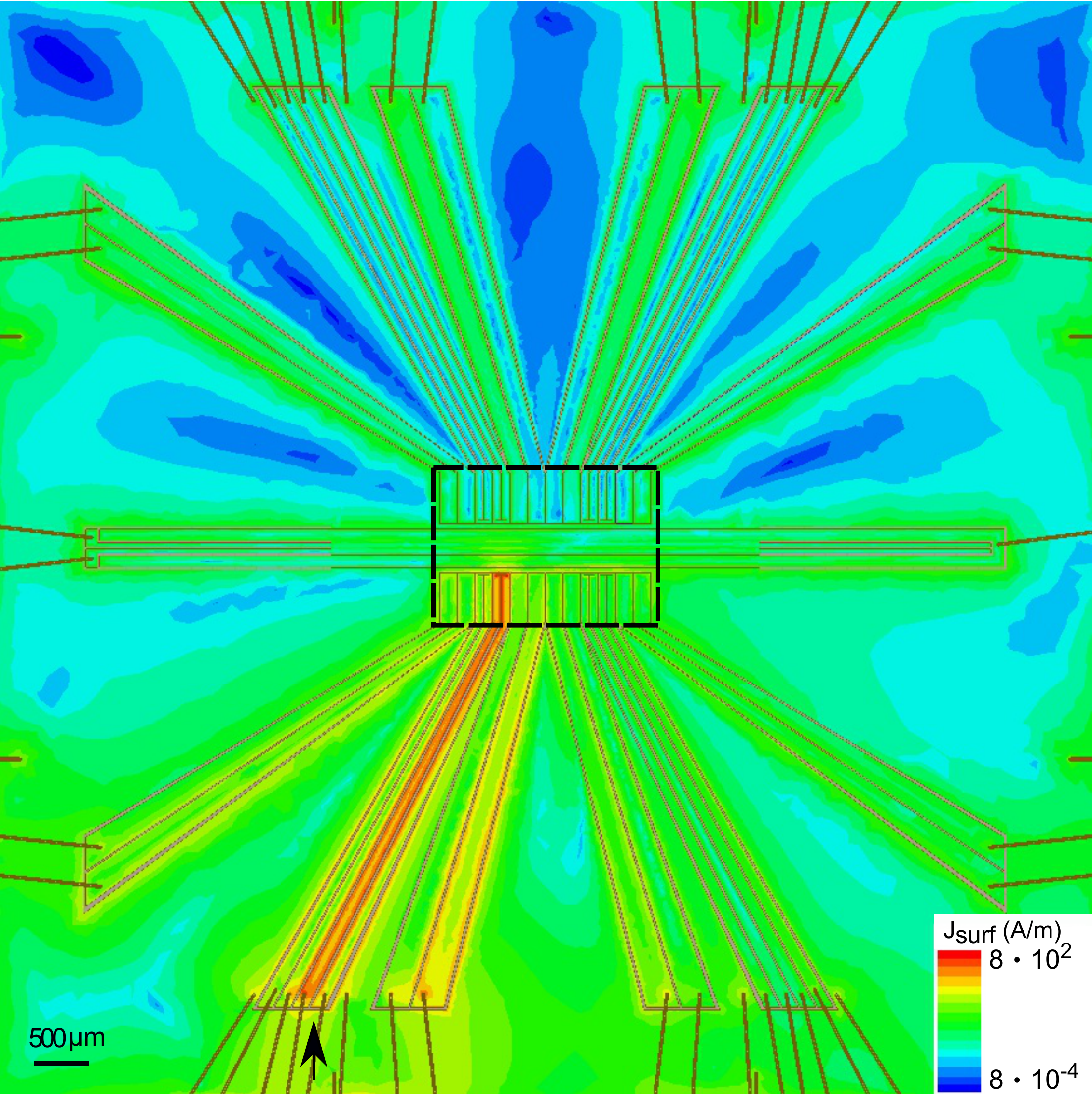}}
\end{center}
\caption{HFSS model of surface current density, $ J_\text{surf}$, on our trap design, when a single microwave control electrode (indicated by arrow) is in use. Induced currents in other electrodes are clearly visible on the log scale used for illustration. See fig.~\ref{nullschematic} for a zoomed-in schematic of the trapping zone (enclosed by the dashed line). }
\label{hfss} 
\end{figure}

The two microwave addressing zones are separated by 960\,$\mu$m.  Two pairs of standard DC electrodes situated either side of the addressing zones permit shuttling of ions between zones.  Away from the centre of the trap, the DC and microwave electrodes are separated by sections of ground plane (see fig.~\ref{hfss}).  This is done to reduce the coupling between the microwave electrodes and the DC electrodes.  Similarly, a strip of ground plane is placed between the microwave and RF electrodes.

\subsubsection{Fabrication and packaging}

The trap is fabricated from electroplated gold on fused silica using the same microfabrication technique used in \cite{Allcock10}.

\begin{figure}
\begin{center}
\resizebox{0.48\textwidth}{!}{\includegraphics{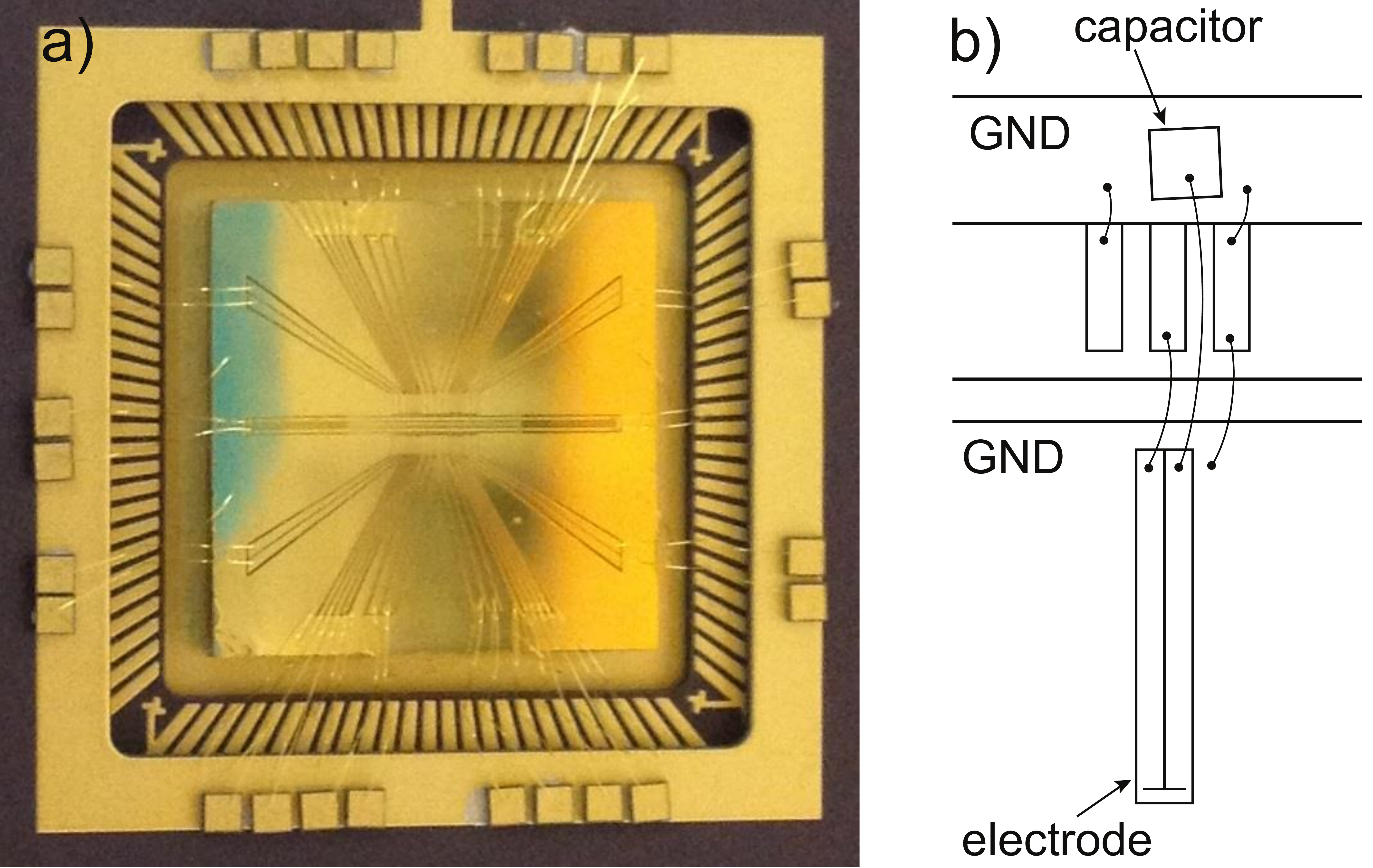}}
\end{center}
\caption{a) The prototype trap on a CPGA package with wirebond connections.  b) Diagram of bondwire connections to a single microwave control electrode.}
\label{package}
\end{figure}

We package the trap in a commercially available 100-pin ceramic pin grid array package (CPGA, see fig.~\ref{package}a).  This package is commonly used in ion-trapping and compatible in-vacuum sockets have already been developed \cite{StiTh}.

Each microwave bond pad connection through the package is made such that there is a grounded bond pad either side (see fig.~\ref{package}b).  The microwaves are brought in on coaxial cable with the centre conductor connected to a microwave pin and the ground split and connected to two ground pins on either side.  This is done to reduce ground discontinuities, which cause reflections, and to reduce crosstalk within the package. 

Both the metallized  bottom of the package cavity and the outer ring around the bond pads are grounded.  The outer ring is used to mount grounding capacitors for all the DC and microwave electrodes (see \cite{Allcock:2011}).  All wire bonds are made with 25\,$\mu$m diameter gold wire.

A test was carried out where a small section of microstrip transmission line was used in place of the trap.  From this, the upper limit on the insertion loss of the package (from coax to trap) was deduced to be between 3.5 and 2.2\,dB in the range $3.1-3.3$\,GHz.

\subsection{Simulated performance}

To simulate the performance of the trap, the $6\times 8$ matrix $\mathbf{M}$ was extracted from the HFSS model. We wish to solve eq. \ref{fieldeqn} to find a set of currents $\mathbf{I_{1}}\ldots\mathbf{I_{8}}$  that need to be applied to the eight electrodes to produce the desired fields at each ion. Because we use eight microwave electrodes to determine six field components (the $x$, $y$ and $z$ components of the fields at the two ions), the problem is under-constrained.  We use the particular solution given by:

\begin{equation}
\label{pinv}
	{\bf I} = \text{pinv}({\bf M}) \cdot {\bf B}
\end{equation}
where $\text{pinv}(\mathbf{M})$  is the Moore-Penrose pseudo-inverse of $\mathbf{M}$. This solution minimizes $  ||{\bf I}|| = \sqrt{|I_1|^2 + \ldots + |I_8|^2 }$.

We are also interested in evaluating the crosstalk when no nulling fields are applied, as this gives a good indication of whether a design flaw is permitting excessive coupling of the microwaves between the two trapping zones. We ask for the magnitude of the ${\bf B}$-field seen by the neighbour when we address an ion using only the four electrodes surrounding it. This is again an under-determined problem, with solutions of the form:

\begin{equation}
\label{xtalksoln}
	{\bf I} = {\bf I_p } + \tilde {c}_n{\bf I_n}
\end{equation}
where $\bf I$ is a 4-component vector that solves eq. \ref{fieldeqn} (with ${\bf M}= {\bf M}_{3\times 4}$ a 3$\times$4 matrix and ${\bf B}$ the vector describing the three components of the ${\bf B}$-field we desire at the addressed ion). $ {\bf I_p} = \text{pinv}({\bf M}_{3\times 4}) \cdot{\bf B}$ is a particular solution, $\tilde {c}_n$ is a complex number and  ${\bf I_n}$ is a vector in the 1 dimensional null space of ${{\bf M}_{3\times 4}}$.

\begin{figure}
\begin{center}
\resizebox{0.48\textwidth}{!}{\includegraphics{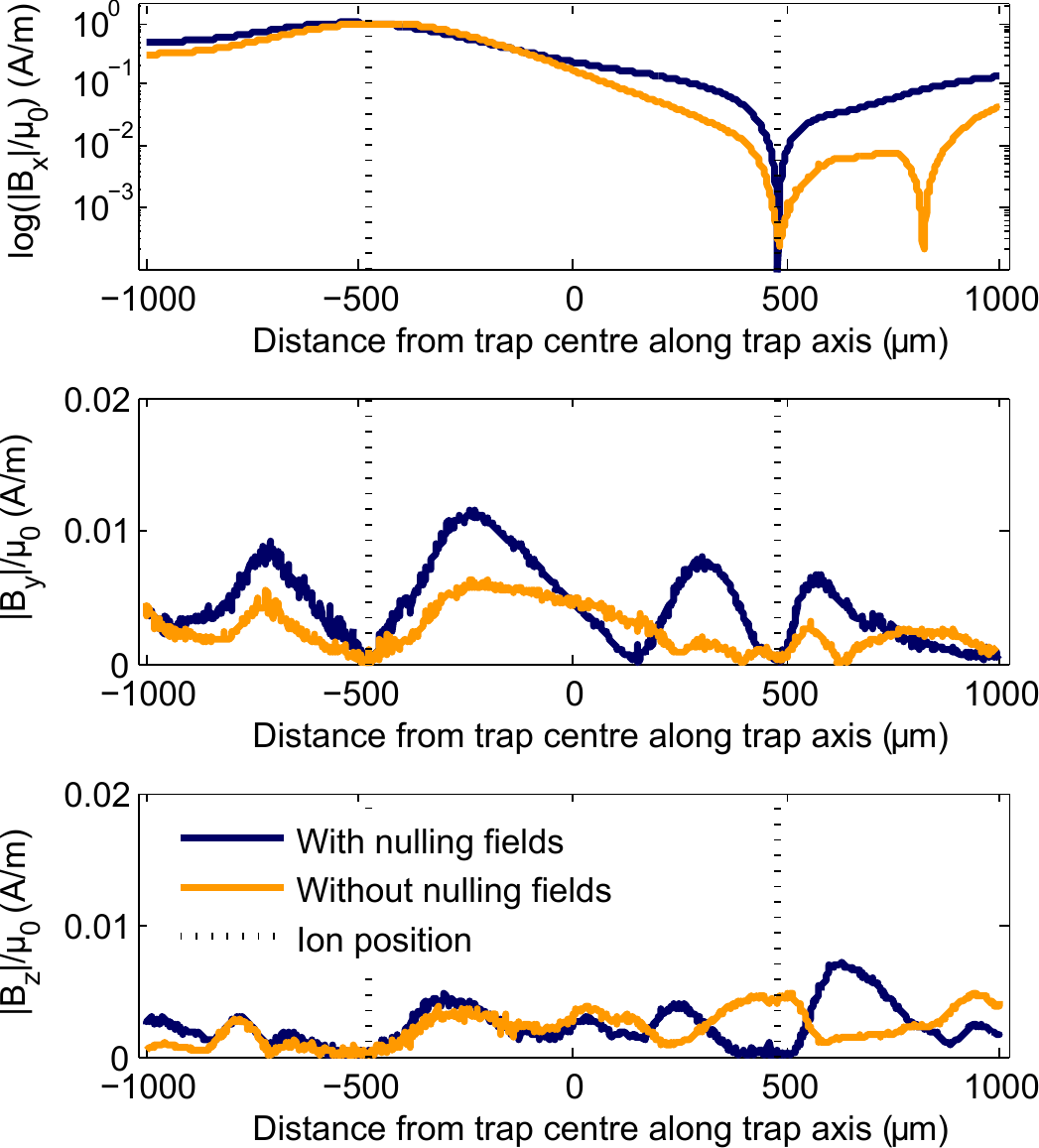}}
\end{center}
\caption{Simulated $x$ (top) , $y$ (middle) and $z$ (bottom) components of the microwave $\bf B$-field amplitude along the trap axis (110 $\mu$m above the electrode surface) when ion 1 is addressed with a field amplitude $B/ \mu_{0} = 1$\,A/m ($B\approx13$\,mG) in the $x$ direction with (dark blue) and without (orange) the use of nulling fields. Note that the plot for the $x$ component is on a log scale.}
\label{fieldstrapaxis} 
\end{figure}

We choose the solution that gives the smallest total field magnitude ($||{\bf B}|| = \sqrt{|B_x|^2 + |B_y|^2+ |B_z|^2 }$) at the neighbour ion; this can be easily found with a constrained search through the null space.  Fig.~\ref{fieldstrapaxis} shows the modelled fields along the trap axis produced when we apply a field $B/ \mu_{0} = 1$\,A/m ($B\approx 13$\,mG) in the $x$ direction to ion 1 with and without the use of nulling fields. We apply the field in the $x$ direction because this is the orientation of the static magnetic field, $\bf B_{0}$, so it is the $x$ component of the microwave field that will drive the low field qubit transition. Even without nulling fields, we see that the crosstalk is $\approx 0.5\%$ .

\begin{figure}
\begin{center}
\resizebox{0.40\textwidth}{!}{\includegraphics{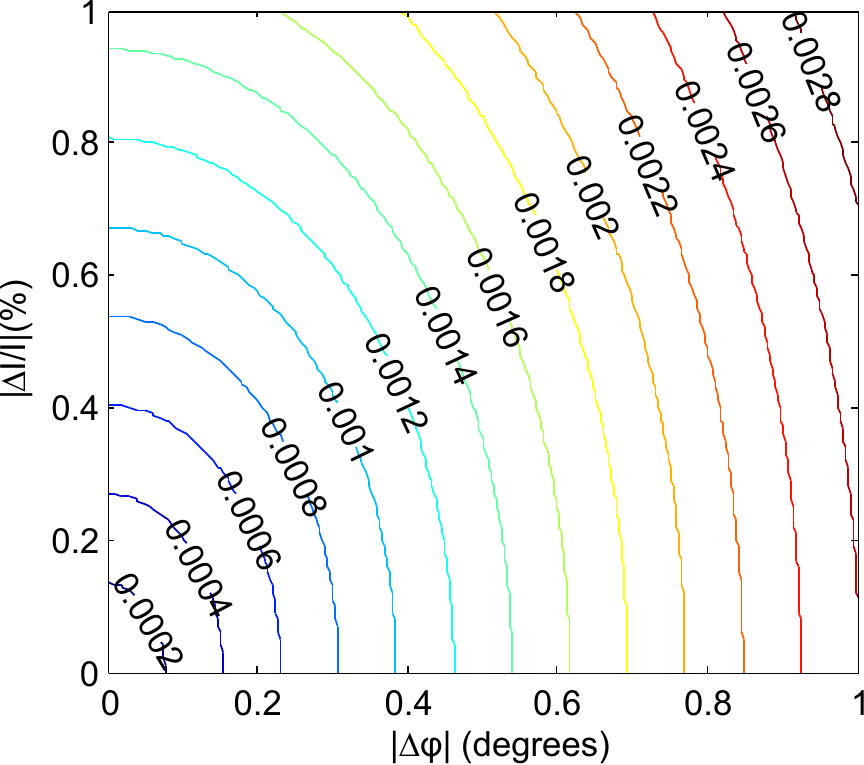}}
\end{center}
\caption{Contour plot of the ratio of the Rabi frequency of the neighbour ion to that of the addressed ion when phase errors $\Delta \phi$ and current errors $\Delta I$ are applied to each of the eight electrodes.}
\label{contour}
\end{figure}

In practice, the level to which we can null the crosstalk will be set by drifts in the current amplitudes and phases we apply to the microwave electrodes (which we expect to dominate over initial calibration errors).  To estimate the change in amplitude of the six ${\bf B}$-field components, $\Delta B_1 , \ldots , \Delta B_6$ , as functions of small fluctuations in current, $\Delta I_1 , \ldots , \Delta I_8$, and phase, $\Delta \phi_1 , \ldots , \Delta \phi_8$, applied to each of the 8 electrodes, we assume that these fluctuations are uncorrelated and add in quadrature: 

\begin{equation}
\label{quadrature}
	\Delta B_j = \sqrt{\sum\limits_{k=1}^8\left| \frac{\partial B_j}{\partial I_k}\right| ^2(\Delta I_k)^2 + \sum\limits_{k=1}^8\left| \frac{\partial B_j}{\partial\phi_k}\right| ^2(\Delta \phi_k)^2 }
\end{equation}where the partial derivatives can be extracted from eq.\ref{fieldeqn} and are given by:

\begin{equation}
\label{errorderivs}
	\frac{\partial B_j}{\partial I_k}  = M_{jk} e^{i\phi_k}\text{\ and\ } \frac{\partial B_j}{\partial\phi_k}  = iM_{jk}I_{k}e^{i\phi_k}
\end{equation}

Fig.~\ref{contour} shows the ratio of the Rabi frequency of ion 2 to that of ion 1 as a function of fluctuations in current and phase on all eight electrodes when ion 1 is addressed by a field in the $x$-direction. For experimentally feasible amplitude and phase stability at the $0.1\%$ and  $0.1\,^{\circ}$ levels respectively, the ratio of the Rabi frequency of the neighbour to that of the addressed ion is ${ \lesssim 0.03\%}$.  We note that techniques such as composite pulse sequences may also be used to improve the fidelity of a given set of gate operations \cite{Bro04}.

Another source of error we consider is spin-motion entanglement due to the microwave field gradient.  From our simulation we calculate the effective Lamb-Dicke parameter (defined as the ratio of the ground state sideband Rabi frequency to the carrier Rabi frequency) to be of order $10^{-5}$.  Therefore, this error will be negligible for this trap.

\subsection{Drive system}

Many channels of phase coherent and easily adjustable microwaves are required for this scheme.  This can be achieved by taking a single microwave local oscillator source, splitting it into as many channels as required and then modulating each channel with a digitally-generated intermediate frequency (IF).  This method provides phase, amplitude and frequency control (the latter only if not using a zero frequency IF).  These techniques are well developed within the telecoms industry and straightforwardly scalable to hundreds of channels.

\section{Future development}

\subsection{Integration with multi-qubit gates}

One of the important advantages of this scheme is that it can also be used to mitigate the crosstalk when carrying out near-field microwave driven multi-qubit entangling gates \cite{Osp08}.  The use of nulling fields allows multi-qubit operations to be performed on a subset of qubits, which are transported to specific multi-qubit gate zones, while other `memory' qubits are left unaffected in the single qubit zones discussed previously.   

The nulling situation here is somewhat different to that previously discussed because the fields required for multi-qubit gates are much larger than those needed for single-qubit operations due to the small effective Lamb-Dicke parameter.  They are, however, detuned from the qubit transition by approximately the secular frequency of the ions in the multi-qubit gate zone (of order $1-10\,$MHz).

These off-resonant fields will introduce a light-shift on the carrier transition of $\Omega^2/2\Delta$, where $\Omega$ is the Rabi frequency they drive and $\Delta$ is the detuning.  Hence, if our nulling reduces the field amplitude by a factor $r$, the light shift will be reduced by $r^2$: we need only null the fields to the 1\% level to reduce the light shift to the 0.01\% level (typically, reducing the Rabi frequency to $<1\,$kHz is sufficient to produce negligible light shift).

There will also be an error associated with the gradient term coupling to the ions' motional sideband transitions and entangling the qubit state with the motional state.  If the ions in the storage zone have a different secular frequency to those in the multi-qubit gate zones (a $>1\,$MHz difference is feasible), these transitions will also be off-resonant and light shifts will similarly be induced.  To estimate the magnitude of these gradients we used the simulation of a multi-qubit gate trap we have in our laboratory \cite{Allcock13} and found that at a distance of  $\sim500\,\mu$m from the multi-qubit gate zone (the ion to electrode distance is 75\,$\mu$m in this trap) the gradient terms were reduced by 2 orders of magnitude.  Therefore, if we are using a ground-state sideband Rabi frequency of  $\sim10$\,kHz in the gate zone, this will be reduced to $<1\,$kHz at the addressing zone, giving a negligible light shift.  If necessary, the gradient term can be reduced further by designing the compensation electrodes such that they can also null the gradient.

\subsection{Scaling up}

The next step is to extend this technique to hundreds of ions in a large array.  In order to do this several challenges will need to be addressed.

\begin{figure}
\begin{center}
\resizebox{0.48\textwidth}{!}{\includegraphics{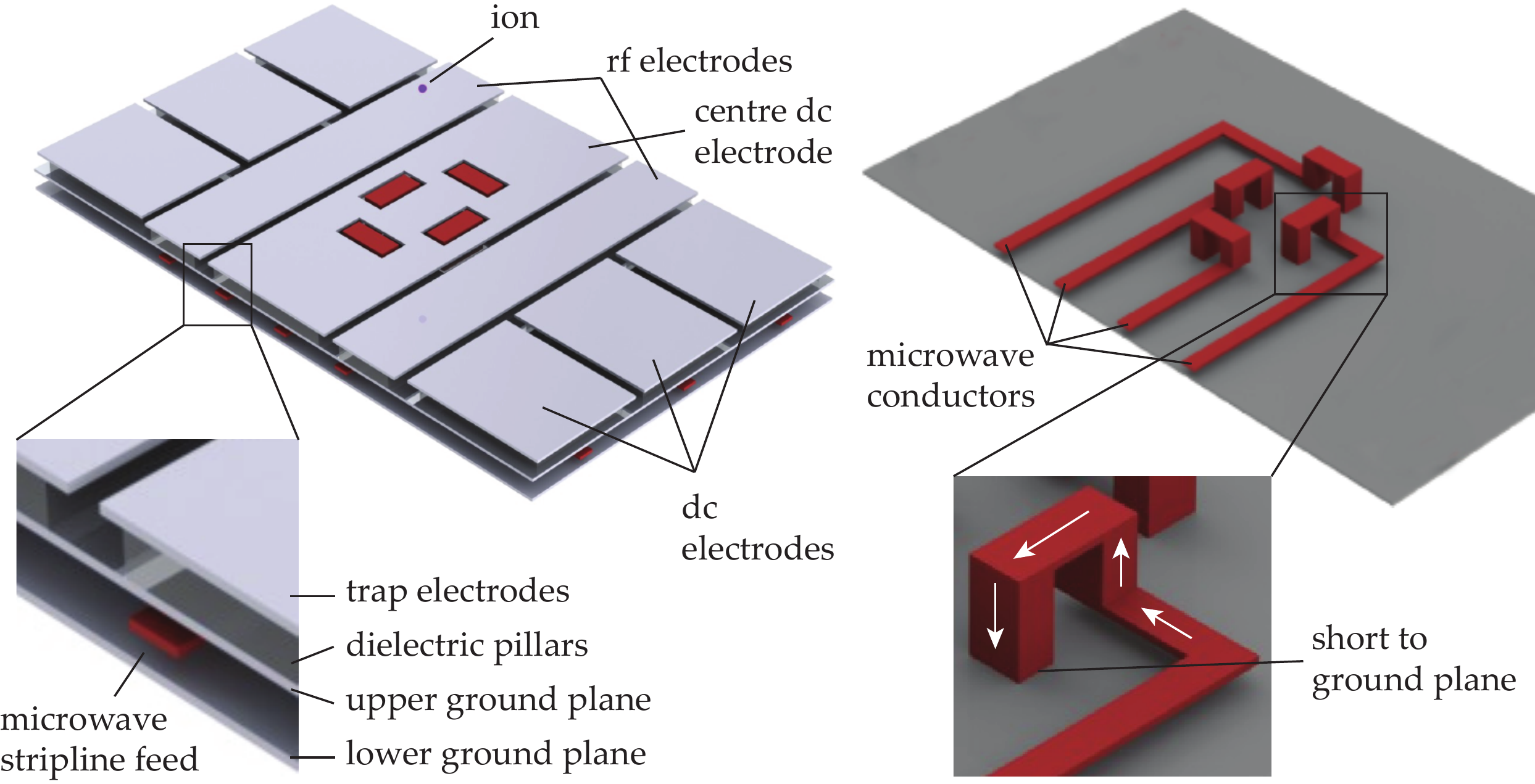}}
\end{center}
\caption{\label{future}An illustration of how a muti-level fabrication technique could allow for a more advanced arrangement of microwave control electrodes and associated feeds.}
\end{figure}

The single-level electrode fabrication technique used here is likely to be insufficient to fabricate a larger array.  Moving to a fabrication technique that allows for multiple dielectric-separated metal levels and vias (for example 3-level metal demonstrated by GTRI \cite{Doret12} and 4-level metal demonstrated by Sandia \cite{SandiaPoster}) would allow for more flexibility.  Firstly, it would allow feed lines to be routed across the trap as stripline between a pair of ground planes (see fig.~\ref{future}), shielding them from other trap zones.  Secondly, if vias are used, the microwave control electrodes can be placed \emph{within} the central DC electrode. This puts them closer to the ion, reducing the microwave power requirement and the crosstalk between zones.  Finally, the integration of capacitors to the ground plane directly under electrodes (as demonstrated in \cite{Allcock:2011}) would allow multi-point grounding of DC electrodes, shorting induced currents to the ground plane and reducing their propagation to other parts of the chip (a significant effect, as can be seen in fig.~\ref{hfss}).  For very large arrays, even this fabrication technology may be insufficient as the number of wirebond interconnects to be made along the edges of the chip will become prohibitive.  One solution may be three-dimensional fabrication techniques such as through-wafer vias \cite{Cho10}.

We also need to consider the scalability of the calibration and nulling procedures.  If we require $m$ measurements to calibrate the effect of one trapping zone on one ion, then, with $N$ ions in the processor (and hence $N$ trapping zones), the calibration of each ion will require $m\cdot N$ measurements. The total number of measurements required to calibrate all $N$ ions is therefore $m\cdot N^2$, which is not a favourable scaling. However, if we can readout all $N$ ions simultaneously, the time it takes to calibrate the processor becomes $t_m = m \cdot N$, which is linear in $N$.  Simultaneous readout can be achieved by imaging the whole processor onto a camera \cite{Bur09} or by having many detectors integrated into the processor, either directly \cite{Eltony13} or coupled via fiber optics \cite{VanDevender10}.

Having more qubits increases the difficulty of nulling the crosstalk since the more crosstalk there is, the more sensitive we are to drifts in phase and frequency. If every additional zone contributed the same crosstalk $\chi_0$, then, for uncorrelated phase and frequency errors, the error from imperfect nulling would increase as $\sqrt{N}$ (by quadrature addition).  However, we must also consider the decrease of the crosstalk with distance $d$: if we assume a power-law decay given by $\frac{1}{d^k}$, zones at a distance $d$ will only contribute crosstalk $\frac{\chi_0}{d^k}$ .  Consider a zone $Z$ at the center of a 2D array of qubit zones spaced by a unit distance.  The number of zones that form a perimeter at distance $d$ from $Z$ is proportional to $d$. If each of these zones contributes a crosstalk of $\frac{\chi_0}{d^k}$, then the perimeter's contribution will be proportional to $\chi_d=\frac{\chi_0}{d^{k-1}}$. If we sum over all distances $d$ we obtain the total crosstalk $\chi_\text{total} = \chi_0 \cdot \sum_{d=1}^{\infty}d^{-k+1}$, which converges for $k>2$. Considering the design illustrated in fig.~\ref{future}, the visible part of each microwave electrode appears at large distance as a small current element: at d.c., this would give a field that decreases as $\sim 1/d^2$ but at microwave frequencies the field will decrease faster because opposing return currents will be induced in the surrounding ground plane. Thus $k>2$ and we expect the crosstalk at any given zone to converge to a fixed value as $N$ increases. We can hence increase the number of qubits in the array indefinitely, whilst maintaining the crosstalk  at any given zone at a fixed value.

\subsection{Conclusion}

We have presented the design of a surface ion trap with integrated microwave electrodes that enables independent frequency, amplitude and phase control of the microwave field at each ion.  Our simulations show that, using nulling fields to eliminate crosstalk in neighbour trapping zones, this design can be used to perform fault-tolerant single-qubit addressing, with crosstalk errors at the 0.01\% level in the presence of realistic microwave phase and amplitude drift. The single-layer design provided for straightforward fabrication of a prototype chip. We discussed the prospect of scaling up the concept using multi-level architectures and, for very large arrays, three-dimensional fabrication techniques. We show that the crosstalk at any given trapping zone due to neighbour zones in a large array converges as the number of zones increases and that the time required to calibrate nulling procedures scales linearly. We propose that it is hence feasible to use this architecture for complete microwave control of parallel single-qubit operations in very large quantum processors. 

\begin{acknowledgement}
We thank D. N. Stacey, H. A. Janacek, M. A. Sepiol, D. Leibfried, J. A. Sherman,  and D. H. Slichter for helpful comments on the manuscript. Thanks to J. Brown and P. Pattinson for the use of cleanroom facilities. This work was supported by an EPSRC Science \& Innovation Award.
\end{acknowledgement}

\bibliographystyle{unsrt}
\bibliography{addpaper}

\end{document}